\documentclass[aps,prd,preprint,tightenlines,nofootinbib]{revtex4}
\usepackage{epsfig}
\usepackage{graphicx}
\usepackage{graphics}
\usepackage{dcolumn}
\usepackage{bm}
\usepackage{afterpage}
\usepackage{citesort}

\newcommand{\etal}{{\it et al.}}

\begin{document}

\preprint{\tighten\vbox{\hbox{\hfil EFI 10-2}
                        \hbox{\hfil SUHEP 10-01}}}

\title
{\LARGE Leptonic Decays of Charged Pseudoscalar Mesons}

\author{Jonathan L. Rosner}
\affiliation{Enrico Fermi Institute, University of Chicago, Chicago, IL 60637}
\author{and Sheldon Stone}
\affiliation{Department of Physics, Syracuse University, Syracuse, NY 13244\\
\\}
\date{\today}

\begin{abstract}
We review the physics of purely leptonic decays of $\pi^\pm$, $K^\pm$,
$D^{\pm}$, $D_s^\pm$, and $B^\pm$ pseudoscalar mesons.  The measured decay
rates are related to the product of the relevant weak-interaction-based CKM
matrix element of the constituent quarks and a strong interaction parameter
related to the overlap of the quark and anti-quark wave-functions in the meson,
called the decay constant $f_P$. The interplay between theory and experiment is
different for each particle. Theoretical predictions of $f_B$ that are needed
in the $B$ sector can be tested by measuring $f_{D^+}$ and $f_{D_s^+}$ in the
charm sector. Currently, these tests are unsatisfactory. The lighter $\pi$ and
$K$ mesons provide stringent comparisons between experiment and theory due to
the accuracy of both the measurements and the theoretical predictions. An
abridged version of this review was prepared for the Particle Data Group's 2010
edition \cite{Previous}.
\end{abstract}
\maketitle

\section{Introduction}
\begin{figure}[hbt]
\centering
\includegraphics[width=3in]{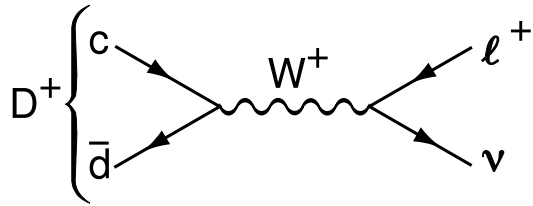}\vskip -0.02mm
\caption{The annihilation process for pure $D^+$ leptonic decays in the
Standard Model.
 } \label{Ptoellnu}
\end{figure}

Charged mesons formed from a quark and anti-quark can decay to a
charged lepton pair when these objects annihilate via a virtual
$W$ boson. Fig.~\ref{Ptoellnu} illustrates this process for the
purely leptonic decay of a $D^+$ meson.

Similar quark-antiquark annihilations via a virtual $W^+$ to
the $\ell^+ {\nu}$ final states occur for the $\pi^+$, $K^+$, $D_s^+$, and
$B^+$ mesons.  (Charge-conjugate particles and decays are implied.) Let $P$ be
any of these pseudoscalar mesons.  To lowest order, the decay width is
\begin{equation}
\Gamma(P\to \ell\nu) = {{G_F^2}\over 8\pi}f_{P}^2\ m_{\ell}^2M_{P}
\left(1-{m_{\ell}^2\over M_{P}^2}\right)^2 \left|V_{q_1
q_2}\right|^2~. \label{equ_rate}
\end{equation}

\noindent Here $M_{P}$ is the $P$ mass, $m_{\ell}$ is the $\ell$
mass, $V_{q_1 q_2}$ is the Cabibbo-Kobayashi-Maskawa (CKM) matrix
element between the constituent quarks $q_1 \bar q_2$ in $P$, and
$G_F$ is the Fermi coupling constant. The parameter $f_P$ is the
decay constant, and is related to the wave-function overlap of the
quark and antiquark.

The decay $P^\pm$ starts with a spin-0 meson, and ends up with a
left-handed neutrino or right-handed antineutrino.  By angular
momentum conservation, the $\ell^\pm$ must then also be left-handed
or right-handed, respectively. In the $m_\ell = 0$ limit, the decay
is forbidden, and can only occur as a result of the finite $\ell$
mass.  This helicity suppression is the origin of the $m_\ell^2$
dependence of the decay width.

There is a complication in measuring purely leptonic decay rates.
The process $P\to \ell\nu\gamma$ is not simply a radiative
correction, although radiative corrections contribute. The $P$ can
make a transition to a virtual $P^*$, emitting a real photon, and
the $P^*$ decays into $\ell\nu$, avoiding helicity suppression. The
importance of this amplitude depends on the decaying particle and
the detection technique.  The $\ell\nu\gamma$ rate for a heavy
particle such as $B$ decaying into a light particle such as a muon
can be larger than the width without photon emission \cite{Bradcor}.
On the other hand, for decays into a $\tau^{\pm}$, the helicity
suppression is mostly broken and these effects appear to be small.

Measurements of purely leptonic decay branching fractions and
lifetimes allow an experimental determination of the product
$\left|V_{q_1 q_2}\right| f_{P}$. If the CKM element is well known
from other measurements, then $f_P$ can be well measured. If, on the
other hand, the CKM element is not well measured, having
theoretical input on $f_P$ can allow a determination of the CKM
element.   The importance of measuring $\Gamma(P\to \ell\nu)$
depends on the particle being considered. For the $B$ system, $f_B$
is crucial for using measurements of $B^0$-$\overline{B}^0$ mixing
to extract information on the fundamental CKM parameters.  Knowledge
of $f_{B_s}$ is also needed, but it cannot be directly
measured as the $B_s$ is neutral, so the violation of the SU(3)
relation $f_{B_s} = f_B$ must be estimated theoretically. This
difficulty does not occur for $D$ mesons as both the $D^+$ and
$D_s^+$ are charged, allowing the direct measurement of SU(3)
breaking and a direct comparison with theory.

For $B^-$ and $D_s^+$ decays, the existence of a charged Higgs boson
(or any other charged object beyond the Standard Model) would modify
the decay rates; however, this would not necessarily be true for
the $D^+$ \cite{Hou,Akeroyd}.  More generally, the ratio of $\tau \nu$
to $\mu \nu$ decays can serve as one probe of lepton universality
\cite{Hou,Hewett}.

\overfullrule 0pt As $|V_{ud}|$ has been quite accurately measured
in super-allowed $\beta$ decays \cite{Vud}, with a value of
0.97425(22) \cite{BM}, measurements of $\Gamma(\pi^+ \to \mu^+{\nu})$ yield a
value for $f_{\pi}$. Similarly, $|V_{us}|$ has been well measured in
semileptonic kaon decays, so a value for $f_{K}$ from $\Gamma(K^-
\to \mu^- \bar{\nu})$ can be compared to theoretical calculations.
Lattice gauge theory calculations, however,  have been
claimed to be very accurate in determining $f_K$, and these have
been used to predict $|V_{us}|$ \cite{Jutt}.

\section{Charmed mesons}
We review current measurements, starting with the charm system.  The CLEO
collaboration has performed the only measurement of the branching fraction for
$D^+\to\mu^+\nu$ \cite{fD}. CLEO uses $e^+e^-$ collisions at the $\psi(3770)$
resonant energy where $D^-D^+$ pairs are copiously produced. They fully
reconstruct one of the $D$'s, find a candidate muon track of opposite sign to
the tag, and then use kinematical constraints to infer the existence of a
missing neutrino and hence the $\mu\nu$ decay of the other $D$. They find
${\cal{B}}(D^+\to\mu^+\nu) = (3.82 \pm 0.32 \pm 0.09) \times 10^{-4}$.
We use the well-measured $D^+$ lifetime of 1.040(7) ps, and assuming $|V_{cd}|$
equals $|V_{us}|=0.2246(12)$ \cite{BM}  minus higher order correction terms
\cite{Charles}, we find $|V_{cd}|=0.2245(12)$.  The CLEO branching fraction
result then translates into a value of
$$
f_{D^+}=(206.7\pm 8.5\pm 2.5)~{\rm MeV}~.
$$
This result includes a 1\% correction (lowering) of the rate due to the
presence of the radiative $\mu^+\nu\gamma$ final state based on the estimate by
Dobrescu and Kronfeld \cite{Kron}.

Before we compare this result with theoretical predictions, we
discuss the $D_s^+$. Measurements of $f_{D_s^+}$ have been made by
several groups and are listed in Table~\ref{tab:fDs}
\cite{CLEO-c, Belle-munu, CLEO-rho, CLEO-CSP, BaBartaunu}.
We exclude values \cite{CLEO, BEAT, ALEPH, L3, OPAL}
 obtained by normalizing to $D_s^+$ decay modes (mentioned in the 2008 version of this Review \cite{Previous}) that are not well defined. Many measurements, for example, used the $\phi\pi^+$ mode. This decay is  a subset of the $D_s^+\to K^+ K^- \pi^+$ channel which has interferences from other modes populating the $K^+K^-$ mass region near the $\phi$, the most prominent of which is the $f_0(980)$. Thus the extraction of effective $\phi\pi^+$ rate is sensitive to the mass resolution of the experiment and the cuts used to define the $\phi$ mass region \cite{reason}\cite{Babar-munu}.

The CLEO and Belle $\mu^+\nu$ results rely on fully
reconstructing all the final state particles except for the neutrino and using
a missing-mass technique to infer the existence of the neutrino. CLEO uses
$e^+e^-\to D_sD_s^*$ collisions at 4170 MeV, while Belle uses $e^+e^-\to D
Kn\pi D_s^*$ collisions at energies near the $\Upsilon(4S)$.

When selecting the
$\tau^+\to\pi^+\bar{\nu}$ and $\tau^+\to\rho^+\bar{\nu}$ decay modes, CLEO uses both
calculation of the missing-mass and the fact that there should be no extra
energy in the event beyond that deposited by the measured tagged $D_s^-$ and the
$\tau^+$ decay products. The $\tau^+\to e^+\nu\bar{\nu}$ mode, however, uses only extra
energy. BaBar measures $\Gamma(D_s+\to \tau^+\nu)/\Gamma(D_s^+\to \overline{K}^0 K^+)$ using the $\tau^+\to e^+\nu \overline{\nu}$ mode. Here the $D_s^-$ tag is formed similarly to Belle by finding events with a $D$ a $K$ and pions opposite a single positron and little extra energy.
Then the analysis is performed selecting modes with a $\overline{K}^0 K^+$ consistent with arising from the decay of a $D_s^+$.
(The fourth error in Table~\ref{tab:fDs} on their measurement reflects the uncertainty due to the PDG value of ${\cal{B}}(D_s^+\to \overline{K}^0K^+)$.)

\begin{table}[htb]
\caption{Experimental results for ${\cal{B}}(D_s^+\to \mu^+\nu)$, ${\cal{B}}
(D_s^+\to \tau^+\nu)$, and $f_{D_s^+}$. Numbers for $f_{D_s^+}$ have been
extracted using updated values for masses and $|V_{cs}|$ (see text); radiative
corrections have been included. Common systematic errors in the CLEO results
have been taken into account. 
\label{tab:fDs}}
\begin{center}
\begin{tabular}{llccc}
 \hline\hline
&Experiment &Mode&${\cal{B}}$&$f_{D_s^+}$ (MeV)\\\hline
&CLEO-c  \cite{CLEO-c}& $\mu^+\nu$& $(5.65\pm
0.45\pm 0.17)\times 10^{-3}$ & $257.6\pm 10.3\pm 4.3$\\
&Belle \cite{Belle-munu}
& $\mu^+\nu$ & $(6.38\pm 0.76\pm 0.57)\times 10^{-3}$& $274\pm 16 \pm 12 $ \\
&Average& $\mu^+\nu$ & $(5.80\pm 0.43)\times 10^{-3}$ & $261.5\pm 9.7$\\
&CLEO-c \cite{CLEO-c}& $\tau^+\nu~(\pi^+\overline{\nu})$ & $(6.42\pm 0.81 \pm
0.18) \times 10^{-2}$& $278.0\pm 17.5 \pm 3.8 $ \\
&CLEO-c \cite{CLEO-rho}& $\tau^+\nu~(\rho^+\overline{\nu})$ & $(5.52\pm 0.57
\pm 0.21)\times 10^{-2}$& $257.8\pm 13.3 \pm 5.2 $ \\
&CLEO-c \cite{CLEO-CSP}& $\tau^+\nu~(e^+\nu\overline{\nu})$ &
$(5.30\pm 0.47\pm 0.22)\times 10^{-2}$& $252.6\pm 11.2 \pm 5.6 $ \\
&BaBar  \cite{BaBartaunu}& $\tau^+\nu~(e^+\nu\overline{\nu})$ &
$(4.54\pm 0.53\pm 0.40\pm 0.28)\times 10^{-2}$ & $233.8\pm 13.7 \pm 12.6 $ \\
&Average & $\tau^+\nu$ & $(5.58\pm 0.35)\times 10^{-2}$ & $255.5\pm 7.5$
\\ \hline
 & Average &$\mu^+\nu$ + $\tau^+\nu$ && $257.5\pm 6.1$\\ \hline \hline
\end{tabular}
\end{center}
\end{table}

We extract the decay constant from the measured branching ratios using the
$D_s^+$ mass of 1.96849(34) GeV,  the $\tau^+$ mass of 1.77684(17) GeV, and a
lifetime of 0.500(7) ps.  We use the first order correction $|V_{cs}| =
|V_{ud}| - |V_{cb}|^2/2$ \cite{Charles} ; taking $|V_{ud}| = 0.97425(22)$
\cite{Vud}, and $|V_{cb}| =0.04$ from an average of exclusive and inclusive
semileptonic $B$  decay results as discussed in Ref.~\cite{Vcb}, we find
$|V_{cs}| = 0.97345(22)$.  Our experimental average,
$$
f_{D_s^+}=(257.5\pm 6.1){\rm ~MeV},
$$
uses only those results that are  included in Table~\ref{tab:fDs}.  We have
included the radiative correction of 1\% in the $\mu^+\nu$ rates listed in the
Table \cite{Kron}~(the $\tau^+\nu$ rates need not be corrected). Other
theoretical calculations show that the $\gamma\mu^+\nu$ rate is a factor of
40--100 below the $\mu^+\nu$ rate for charm \cite{theories-rad}.

\begin{table}[htb]
\caption{Theoretical predictions of $f_{D^+_s}$, $f_{D^+}$, and
$f_{D_s^+}/f_{D^+}$. QL indicates a quenched-lattice calculation, while
PQL indicates a partially-quenched lattice calculation.
(Only selected results having errors are included.)
\label{tab:Models}}
\begin{center}
\begin{tabular}{llccc}
\hline\hline
& Model & $f_{D_s^+}$(MeV) & $f_{D^+}$(MeV) & $f_{D_s^+}/f_{D^+}$\\\hline
& Experiment (our averages)& $257.5\pm 6.1$ &
$206.7\pm 8.9$& $1.25\pm 0.06$ \\ \hline
 & Lattice(HPQCD+UKQCD) \cite{Lat:Foll} & $241\pm 3$& $208\pm 4$& $1.162\pm
0.009$\\ & Lattice (FNAL+MILC+HPQCD) \cite{Lat:Milc} & $260\pm 10$& $217\pm
   10$& $1.20\pm 0.02$\\
&PQL \cite{Lat:Nf2}& $244\pm 8$&$197\pm 9$&
   $1.24\pm 0.03$\\
& QL (QCDSF) \cite{QCDSF} & $220\pm 6\pm 5\pm 11$& $206\pm 6\pm 3\pm 22$&
   $1.07\pm 0.02\pm 0.02$\\
& QL (Taiwan) \cite{Lat:Taiwan} & $266\pm 10\pm 18$& $235\pm 8\pm 14$&
   $1.13\pm 0.03\pm 0.05$\\
& QL (UKQCD) \cite{Lat:UKQCD} & $236\pm 8^{+17}_{-14}$& $210\pm 10^{+17}_{-16}$
   & $1.13\pm 0.02^{+0.04}_{-0.02}$\\
& QL \cite{Lat:Damir} & $231\pm 12^{+6}_{-1}$& $211\pm 14^{+2}_{-12}$&
   $1.10\pm 0.02$\\
& QCD Sum Rules \cite{Bordes} & $205\pm 22$& $177\pm 21$& $1.16\pm 0.01\pm0.03$
\\
& QCD Sum Rules \cite{Chiral} & $235\pm 24$& $203\pm 20$& $1.15\pm 0.04$\\
& Field Correlators \cite{Field} & $260\pm 10$& $210\pm 10$& $1.24\pm 0.03$\\
&Light Front \cite{LF} & $268.3\pm 19.1$ & 206 (fixed) & $1.30\pm 0.04$\\
\hline\hline
\end{tabular}
\end{center}
\end{table}

Two ratios are of particular interest. The ratio of decay constants for the
$\tau^+\nu:\mu^+\nu$ modes is $ f_{D_s^+}(\tau^+\nu)/f_{D_s^+}(\mu^+\nu) =0.98 \pm
0.05$, and the ratio of $D_s^+$ to $D^+$ decay constants is
$f_{D_s^+}/f_{D^+}=1.25\pm 0.06$.

Table~\ref{tab:Models} compares the experimental $f_{D_s^+}$ with theoretical
calculations \cite{Lat:Foll,Lat:Milc,Lat:Nf2,QCDSF,Lat:Taiwan,Lat:UKQCD,%
Lat:Damir,Bordes,Chiral,Field,LF}.  While most theories give values lower than
the $f_{D_s^+}$ measurement, the errors are sufficiently large, in most cases,
to declare success. An unquenched lattice calculation \cite{Lat:Foll}, however,
differs by 2.4 standard deviations \cite{Crit-follana}. Remarkably it agrees
with $f_{D^+}$ and consequently disagrees in the ratio $f_{D_s^+}/f_{D^+}$,
with less significance as the error in $f_{D^+}$ is substantial.

The Fermilab-MILC result has been updated; the preliminary values for $f_{D^+}$
and $f_{D_s^+}$ were raised by 10 MeV and 11 MeV, respectively \cite{Lat:Milc}.
These changes bring the predictions for both numbers within errors of
experiment.

Upper limits on $f_{D^+}$  and $f_{D_s}$ of 230 and 270 MeV, respectively, have
been determined using two-point correlation functions by Khodjamirian
\cite{Kho}.  The $D^+$ result is safely below this limit, while the average
$D_s$ result is also, but older results \cite{Previous} not used in our average
are often above the limit.

Akeroyd and Chen \cite{AkeroydC} pointed out that leptonic decay widths are
modified in two-Higgs-doublet models (2HDM).  Specifically, for the $D^+$ and
$D^+_s$, Eq.~\ref{equ_rate} is modified by a factor $r_q$ multiplying the
right-hand side \cite{AkeroydM}:


$$
r_q=\left[1+\left(1\over{m_c+m_q}\right)\left({M_{D_q}\over M_{H^+}}\right)^2
\left(m_c-\frac{m_q\tan^2\beta}{1+\epsilon_0\tan\beta}\right)\right]^2,
$$

\noindent where $m_{H^+}$ is the charged Higgs mass, $M_{D_q}$ is
the mass of the $D$ meson (containing the light quark $q$), $m_c$ is
the charm quark mass, $m_q$ is the light-quark mass, and $\tan\beta$
is the ratio of the vacuum expectation values of the two Higgs
doublets. In models where the fermion mass arises from coupling to more
than one vacuum expectation value $\epsilon_0$ can be non-zero, perhaps
as large as 0.01. For the $D^+$, $m_d
\ll m_c$, and the change due to the $H^+$ is very small. For the
$D_s^+$, however, the effect can be substantial.

A major concern is the need for the Standrd Model (SM) value of $f_{D_s^+}$.
We can take that from a theoretical model. Our most aggressive choice is that
of the unquenched lattice calculation \cite{Lat:Foll}, because they claim the
smallest error.  Since the charged Higgs would lower the rate compared to the
SM, in principle, experiment gives a lower limit on the charged Higgs mass.
However, the value for the predicted decay constant using this model is 2.4
standard deviations {\it below} the measurement, implying that (a) either the
model of Ref.~\cite{Lat:Foll} is not representative; (b) no value of $m_{H^+}$
in the two-Higgs doublet model will satisfy the constraint at 99\% confidence
level; or (c) there is new physics, different from the 2HDM, that interferes
constructively with the SM amplitude such as in the R-parity-violating model of
Akeroyd and Recksiegel \cite{Rviolating}. Also in the context of R parity
violation, Kundu and Nandi relate this discrepancy with preliminary indications
of a large phase in $B_s-\overline{B_s}$ mixing and explain both with a
specific supersymmetry model \cite{KN}.

Dobrescu and Kronfeld \cite{Kron} emphasize that the discrepancy between the
theoretical lattice calculation and the CLEO data is substantial and ``is worth
interpreting in terms of new physics'' (at least prior to the change in
Fermilab-MILC result and the updated experimental values).  They give three
possible examples of new physics models that might be responsible.  These
include two leptoquark models, and a specific two-Higgs doublet model which
leads to constructive interference with the Standard Model where one doublet
gives the $c$ and $u$ quark masses and the lepton masses, but not the
$d,~ s,~ b,$  or $t$ masses.
(However, for constraints on models of R-parity-violating supersymmetry and
leptoquarks, see Refs.\ \cite{Dorsner:2009}.) 
Gninenko and Gorbunov argue that the neutrino in
the $D_s$ decay mixes with a sterile neutrino, which enhances the rate and also
explains the excess number of low energy electron-like events in the MiniBooNE
data \cite{GG}.

Akeroyd and Mahmoudi \cite{AkeroydM} point out that new physics can affect
the $\mu^+\nu$ and $\tau^+\nu$ final states differently and thus should be
studied separately. They present constraints for the charged Higgs mass in a
specific SUSY model, the Non-Universal Higgs Mass, using the branching ratios
values for $D_s^+\to\tau^+\nu$ and $\mu^+\nu$ separately. These constraints in
some regions are better than any other. The model of Gninenko and Gorbunov is
an example of a model where $\mu^+\nu$ and $\tau^+\nu$ should be treated
separately in that a sterile neutrino is likely to couple very differently to
$\nu_{\mu}$ and  $\nu_{\tau}$.  Other theoretical papers that are pertinent to
this discussion concern leptoquark models \cite{Benbrik}, R-parity-violating
models \cite{Dey,BCN}, 2HDM \cite{Logan}, anomalous $W$-boson charm
quark couplings \cite{HTV}, unparticle physics \cite{WKY}, and constraints on
2HDM models \cite{MO,Deschamps}.

To sum up, the situation is not clear. To set limits on new physics we need an
accurate calculation of $f_{Ds}$ and more precise measurements would also be
useful.

\section{\boldmath\bf The $B$ meson}
The Belle and BaBar collaborations have found evidence for $B^-\to\tau^-
\bar{\nu}$ decay in $e^+e^-\to B^-B^+$ collisions at the $\Upsilon(4S)$ energy.
The analysis relies on reconstructing a hadronic or semi-leptonic $B$ decay
tag, finding a $\tau$ candidate in the remaining track and or photon
candidates,  and examining the extra energy in the event which should be close
to zero for a real $\tau$ decay opposite a $B$ tag. The results are listed in
Table~\ref{tab:Btotaunu}.

\begin{table}[htb]
\caption{Experimental results for ${\cal{B}}(B^-\to \tau^-\overline{\nu})$.
We have computed an average for the two Belle measurements assuming that the
systematic errors are fully correlated.\label{tab:Btotaunu}}
\begin{center}
\begin{tabular}{lllc} \hline\hline
&Experiment & Tag &${\cal{B}}\times 10^{-4} $\hfill\\
\hline
&Belle~\cite{BelleH}&Hadronic&$(1.79^{+0.56~+0.46}_{-0.49~-0.51})$\\
&Belle~\cite{BelleS}&Semileptonic&$(1.65^{+0.38~+0.35}_{-0.37~-0.34})$\\
&Belle&Our Average&$(1.70^{+0.47}_{-0.46})$\\
&BaBar~\cite{BaBarH}&Hadronic&$(1.8^{+0.9}_{-0.8}\pm 0.4)$\\
&BaBar~\cite{BaBarS}&Semileptonic&$(1.7\pm 0.8\pm 0.4)$\\
&BaBar~\cite{BaBarS}&Average&$(1.8^{+1.0}_{-0.9})$\\\hline
&&Our Average&$(1.72^{+0.43}_{-0.42})$\\
\hline\hline
\end{tabular}
\end{center}
\end{table}

There are large backgrounds under the signals in all cases. The systematic
errors are also quite large, on the order of 20\%. Thus, the significances are
not that large.  Belle quotes 3.5$\sigma$ and 3.8$\sigma$ for their hadronic
and semileptonic tags, respectively, while BaBar quotes 2.8$\sigma$ for their
combined result. We note that the four central values are remarkably close to
the average considering the large errors on on all the measurements.
More accuracy would be useful to investigate the effects of new physics.
Here the effect of a charged Higgs is different as it can either increase or
decrease the expected SM branching ratio. The factor $r$ in the 2HDM that
multiplies the right side of Eq.~\ref{equ_rate} is given in terms of the $B$
meson mass, $M_B$, by \cite{Hou,AkeroydM}
\begin{equation}
r=\left(1-\frac{\tan^2\beta}{1-\epsilon_0\tan\beta} {M_B^2\over m^2_{H^+}}\right)^2.
\label{eq:Hou}
\end{equation}

We can derive limits in the 2HDM  $\tan\beta$--$m_{H^+}$ plane. Again, we need to
know the SM prediction of this decay rate. We ascertain this value
using Eq.~\ref{equ_rate}. Here theory provides a value of
$f_B=(193\pm 11)$ MeV \cite{fBl}.
We also need a value for $|V_{ub}|$. Here significant differences
arise between using inclusive charmless semileptonic decays and
the exclusive decay $B\to\pi\ell^+\nu$ \cite{ABS}.  We find that
the inclusive decays give rise to a value of $|V_{ub}|=(4.21\pm 0.25)\times 10^{-3}$,
while the $\pi\ell^+\nu$ measurements yield
$|V_{ub}|=(3.50\pm 0.35)\times 10^{-3}$.
Taking an average
over inclusive and exclusive determinations, and enlarging the error
using the PDG prescription because the results differ, we find
$|V_{ub}|=(3.97\pm0.55)\times 10^{-3}$, where the error is dominantly
theoretical. We thus arrive at the SM prediction for the $\tau^-\bar{\nu}$
branching fraction of $(1.04\pm 0.31)\times 10^{-4}$.

Taking the ratio
of the experimental value to the predicted branching ratio at its
90\% c.l.\ {\it upper} limit and using
Eq.~\ref{eq:Hou} with $\epsilon_0$ set to zero, we find
that we can limit $M_{H^+}~/\tan \beta > 3.3$ GeV.  The
90\% c.l.\ {\it lower} limit also permits us to exclude the region
3.8 GeV $< M_{H^+}~/\tan \beta < 18.0$ GeV \cite{IsiPar}. Considering
the large uncertainties on $V_{ub}$ and the branching ratio measurements, this should be taken more as indication of what the data can eventually tell us when and if the situation improves.

\section{Charged pions and kaons}
We now discuss the determination of charged pion and kaon decay constants.
The sum of branching fractions for
$\pi^- \to \mu^- \bar \nu$ and $\pi^- \to \mu^- \bar \nu \gamma$ is
99.98770(4)\%.  The two modes are difficult to separate experimentally, so we
use this sum, with Eq.~\ref{equ_rate} modified to include photon emission
and radiative corrections \cite{Marciano-Sirlin}.  The branching fraction together
with the lifetime 26.033(5) ns gives
$$
 f_{\pi^-} = (130.41\pm 0.03\pm 0.20)~{\rm MeV}~.
$$
\noindent The first error is due to the error on $|V_{ud}|$,
0.97425(22) \cite{Vud}; the second is due to the higher-order
corrections, and is much larger.

Similarly, the sum of branching fractions for $K^- \to \mu^- \bar
\nu$ and $K^- \to \mu^- \bar \nu \gamma$ is 63.55(11)\%, and the
lifetime is 12.3840(193) ns \cite{Flavi}.
Measurements of semileptonic kaon decays
provide a value for the product $f_+(0)|V_{us}|$, where $f_+(0)$ is the form-factor at
zero four-momentum transfer between the initial state kaon and the final state pion. We use a value for
$f_+(0)|V_{us}|$
of 0.21664(48) \cite{Flavi}. The $f_+(0)$ must be determined theoretically. We
follow Blucher and Marciano \cite{BM} in using the lattice calculation
$f_+(0)=0.9644\pm 0.0049$ \cite{latticefp}, since it appears to be
more precise than the classic Leutwyler-Roos calculation
$f_+(0)=0.961\pm 0.008$ \cite{LR}. The result is $|V_{us}|=0.2246\pm 0.0012$, which is
consistent with the hyperon decay value of $0.2250\pm 0.0027$ \cite{hyperon}.
We derive
$$
f_{K^-} =(156.1 \pm 0.2\pm 0.8\pm 0.2)~{\rm MeV}~.
$$

\noindent The first error is due to the error on $\Gamma$; the
second is due to the CKM factor $|V_{us}|$, and the third is due to the higher-order
corrections. The largest source of error in these corrections
depends on the QCD part, which is based on one calculation in the
large $N_c$ framework.  We have doubled the quoted error here; this
would probably be unnecessary if other calculations were to come to
similar conclusions.  A large part of the additional uncertainty
vanishes in the ratio of the $K^-$ and $\pi^-$ decay constants,
which is
$$
f_{K^-}/f_{\pi^-} = 1.197 \pm 0.002 \pm 0.006 \pm 0.001~.
$$

\noindent
The first error is due to the measured decay rates; the second is due to the
uncertainties on the CKM factors; the third is due to the uncertainties in the
radiative correction ratio.

These measurements have been used in conjunction with calculations of
$f_K/f_{\pi}$ in order to find a value for $|V_{us}|/|V_{ud}|$.  Two recent
lattice predictions of $f_K/f_{\pi}$ are $1.189 \pm 0.007$ \cite{Lat:Foll}
and $1.192 \pm 0.007 \pm 0.006$ \cite{fkfpi}.  Together with the precisely
measured $|V_{ud}|$, this gives an independent measure of $|V_{us}|$
\cite{Jutt,Flavi}.

We gratefully acknowledge support of the U. S. National Science Foundation
and the U. S. Department of Energy through Grant No.\ DE-FG02-90ER40560.

\end{document}